# Multi-resonant silver nano-disk patterned thin film hydrogenated amorphous silicon solar cells for SWE compensation

Ankit Vora,[1] Jephias Gwamuri,[2] Joshua M. Pearce,[1,2] Paul L. Bergstrom,[1] and Durdu Ö. Güney[1,a]

[1] *Department of Electrical and Computer Engineering, Michigan Technological University, Houghton, MI, 49931, USA*

[2] *Department of Materials Science and Engineering, Michigan Technological University, Houghton, MI, 49931, USA*

We study polarization independent improved light trapping in commercial thin film hydrogenated amorphous silicon (a-Si:H) solar photovoltaic cells using a three-dimensional silver array of multi-resonant nano-disk structures embedded in a silicon nitride anti-reflection coating (ARC) to enhance optical absorption in the intrinsic layer (i-a-Si:H) for the visible spectrum for any polarization angle. Predicted total optical enhancement (OE) in absorption in the i-a-Si:H for AM-1.5 solar spectrum is 18.51% as compared to the reference, and producing a 19.65% improvement in short-circuit current density ($J_{SC}$) over 11.7 mA/cm$^2$ for a reference cell. The $J_{SC}$ in the nano-disk patterned solar cell (NDPSC) was found to be higher than the commercial reference structure for any incident angle. The NDPSC has a multi-resonant optical response for the visible spectrum and the associated mechanism for OE in i-a-Si:H layer is excitation of Fabry-Perot resonance facilitated by surface plasmon resonances. The detrimental Staebler-Wronski effect (SWE) in a-Si:H solar cell can be minimized by the additional OE in the NDPSC and self-annealing of defect states by additional heat generation, thus likely improving the overall stabilized characteristics of a-Si:H solar cells.

## I. INTRODUCTION

Thin-film hydrogenated amorphous silicon (a-Si:H) solar photovoltaic (PV) cells have the fastest energy payback time of any Si-based PV cell.[1] Amorphous silicon thin-film solar cells with conversion efficiencies over 10% have been reported,[2] however there is an increased interest to further improve the efficiency and reduce the cost simultaneously for broader commercialization and lower levelized costs of solar electricity.[3] The greatest technological challenge encountered by a-Si:H PV is light-induced degradation of performance known as the Staebler-Wronski effect (SWE),[4] which is reversible with thermal annealing. SWE is associated with the formation of defect states in the bandgap from exposure to sunlight, which causes a decrease in a-Si:H PV conversion efficiency.[5] It is now clear that SWE is caused by an increased density of multiple types of defect states, which reach saturation in device quality materials known as the degraded steady state (DSS) after approximately 100 hours of 1 sun light illumination.[6-8] SWE has been studied in detail for decades and several engineering techniques have been used to minimize its impact including various forms of optical enhancement[9] (OE) and even in-situ annealing in photovoltaic thermal hybrid systems,[10, 11] but it has not been eliminated. Therefore, SWE limits the thickness of the intrinsic (i-a-Si:H) layer, and hence the overall absorption capacity of an a-Si:H PV cell.

---

[a] Author to whom correspondence should be addressed. E-mail: dguney@mtu.edu

Recent advances in plasmonics provide a new technique to improve the optical enhancement in a-Si:H PV devices and further reduce the negative effects of SWE. Employing plasmonic nanostructures in PV is garnering broad interest[12] because resonant plasmonic nanostructures are capable of producing optical enhancement in absorption by supporting mechanism like Fabry-Perot resonance, guided modes, localized surface plasmon resonance (LSPR) or increased scattering. Their application has also been studied for PV design, which can result in broadband, polarization independent and wide angle absorption for ultrathin active absorbing layers (<100 nm).[13-21] Three major geometries with associated enhancement mechanisms have been proposed: (i) plasmonic nanostructures on top, (ii) plasmonics nanostructures embedded into the active layer or (iii) textured back contact.[22, 23] For the latter, it has been reported that depositing a-Si:H on such textured or patterned surfaces results in higher defects density and hence reduces open circuit voltage ($V_{oc}$), short circuit current density ($J_{sc}$) and thus the conversion efficiency.[24, 25] Therefore, top surface texturing or front nano-patterned resonant metallic nanostructures coupled with the active semiconductor seems to be a more promising option for optical absorption enhancement in the active layer while minimizing the defect density and at the same time facilitating the feasibility of commercial production of high efficiency a-Si:H PV devices. Massiot et al.[21] have theoretically demonstrated broadband absorption in their ultrathin (<100 nm) a-Si:H solar cell using sub-wavelength nano-patterned horizontal nanowires, which also managed to reduce the effect of SWE in a-Si:H. However, their design presents a number of practical limitations. Firstly, plasmonic enhancement is a near field enhancement and, therefore, the strong electric field enhancement due to surface plasmons or guided modes tends to be strongly localized near the metal/semiconductor interface (or even the buffer layer). In the case of commercial a-Si:H cells, it is the top p-a-Si:H and contact layer made from indium tin oxide (ITO), which absorbs enhanced near field more strongly.[21] Secondly, the effects of absorption in the highly defective doped regions of p and n-a-Si:H layers are also not considered in most of the available literature. This is a fundamental omission, since the p-i-n device was developed so that the relatively low-defect density i-layer minimizes losses from the electron-hole pair generation in the higher defect density doped layers.[26] These doped layers have a substantial thickness by surface plasmonic standards as they are usually 15-25 nm thick, and hence can account for a significant amount of the 'enhanced' optical absorption while not making a major contribution to carrier collection. The ability of device designers to compress the doped layer thicknesses further is limited by the design requirement to generate a suitable electric field to separate the solar photo-generated charge carriers. The optical absorption in the p or n a-Si:H layer is largely converted into heat via recombination of electron hole pairs rather than charge carrier separation.[27]

In this paper, a design approach is presented to improve the performance characteristics of a-Si:H commercial solar cells by employing the advantages of resonant plasmonic nano-structures and also combat the negative effects of SWE. We have taken into account the effects of absorption in p and n a-Si:H layers of a commercial structure and applied the technique of optical enhancement in i-a-Si:H layer using resonant plasmonic nano-structures without altering the parameters of the semiconductor layers. We have studied the effect of introducing nano-patterned metallic structures on the absorption characteristics of commercial a-Si:H solar cell (for each individual layer) and also discussed the associated physical mechanism for optical enhancement.

**II. METHODS**

The complex refractive index of p-a-Si:H, i-a-Si:H, n-a-Si:H, and aluminum doped zinc oxide (AZO) layers of commercial a-Si:H solar cell (fabricated by ThinSilicon, Mountain View, CA) were measured using a J.A. Woollam variable-angle spectroscopic ellipsometer, Fig. 1(a). The refractive index of both silver and silicon nitride ($Si_3N_4$), which functions as an anti-reflection coating (ARC), were taken from[28] and ITO was taken from SOPRA database.[29] Optical responses of the reference cell and nano-disk patterned solar cell (NDPSC) were obtained through a fully-vectorial finite element based commercial software package COMSOL Multiphysics RF module v4.3b in frequency domain coupled with MATLAB R2012a. For normal incidence response, both reference and NDPSC solar cells were simulated using periodic boundary conditions for the vertical boundaries due to periodic arrangement of unit cell and were excited (excitation port) from air above the ARC (ITO for reference solar cell) and output port was set in air below the silver ground plate. The absorbance in individual layers was calculated from the power loss density function in COMSOL, and this absorbance in i-a-Si:H layer of solar cells was used to calculate theoretical absorbed power density in NDPSC and reference structure, for incident AM 1.5 reference solar spectrum using the equation:

$$P_{i-a-Si:H} = \int A(\lambda) E_{AM1.5}(\lambda) d\lambda \qquad (1)$$

where $P_{i-a-Si:H}$ is the absorbed power density (measured in W/m$^2$) in the i-a-Si:H layer for AM 1.5 reference solar spectrum, $A(\lambda)$ is absorbance (i.e. the ratio of total absorbed power to input optical power) in i-a-Si:H layer as a function of wavelength, and $E_{AM1.5}(\lambda)$ is the spectral irradiance as a function of wavelength (obtained from NREL[30]). The optical enhancement (OE) is defined and calculated using the expression:

$$OE = (P_{i-a-Si:H(NDPSC)} / P_{i-a-Si:H(Ref)} - 1) \times 100 \qquad (2)$$

where the subscript *'NDPSC'* and *'Ref'* denote NDPSC solar cell and reference cell, respectively. Assuming that all photo-generated carriers are collected, the theoretical maximum short-circuit current density ($J_{SC}$) can be calculated using the expression:

$$J_{sc} = \int \frac{q}{hc} A(\lambda)\lambda E_{AM1.5}(\lambda)d\lambda \tag{3}$$

where $q$ is the electron charge, $c$ is the speed of light in vacuum, $h$ is the Planck constant, and $E_{AM1.5}(\lambda)$ is the spectral irradiance. The oblique incidence response of both reference and NDPSC was modeled for their respective structures using periodic boundary condition with Floquet periodicity for incidence angle up to 80° with normal.

Individual layers of the reference cell are depicted in Fig. 1(b), the top layer is an ITO of 70 nm thickness, which is a lossy contact layer, also serving as an ARC. Then p-a-Si:H is 17.5 nm thick, followed by i-a-Si:H which is 350.5 nm and n-a-Si:H as 22.5 nm, AZO of 100 nm thickness and the back silver reflector is 200 nm thick. The optical response of reference structure for normal incidence is shown in Fig. 1(c).

## III. RESULTS AND DISCUSSION

We are presenting a design approach to maximize the absorption in i-a-Si:H layer of commercial a-Si:H solar cell (at normal and oblique incidence) and simultaneously reduce the effects of undesirable SWE by replacing the top contact layer (ITO) with an array of silver nano-disk patterned structures embedded into a silicon nitride ARC. This particular geometry of nano-disks was selected due to its top structure axial isotropy that leads to a polarization independent optical response. Other sub-wavelength geometry were explored such as a 1-D nanowire array, a 2-D grid structure, and nano-sphere patterns, but the nano-disk geometry worked most effectively to increase absorption and provide a polarization independent response. In this paper, this concept is presented using silver nano-disks of diameter 240 nm and height 50 nm, embedded into silicon nitride ARC of thickness 60 nm, as shown in Fig. 2(a). The period of the unit cell is 550 nm and a 10 nm thick ITO layer serves as a buffer layer to stop the diffusion of silver into the p-a-Si:H. The ITO layer also aids in fine tuning the resonance towards shorter wavelengths. This NDPSC design has feature parameters that are scalable for commercial viability using fabrication techniques like nano-imprint lithography[31] or extreme ultraviolet lithography suitable for subwavelength size features for plasmonic applications in solar cells.[12] It was found that the silicon nitride ARC layer performed best by maximizing absorption when it was 60 nm thick, which is consistent with Massiot, et al.[21] The ITO layer was preferred to be kept 10 nm in thickness such that it would minimize the parasitic Ohmic losses, simultaneously fulfill the purpose as a buffer layer, and to tune the resonance towards a blue shift. The optical response of NDPSC for all layers at normal incidence is displayed in Fig. 2(b) with absorbance in the i-a-Si:H reference for comparison. The short circuit current density for both reference structure and NDPSC at normal and oblique incidence, calculated using equation (3) is plotted in Fig. 3.

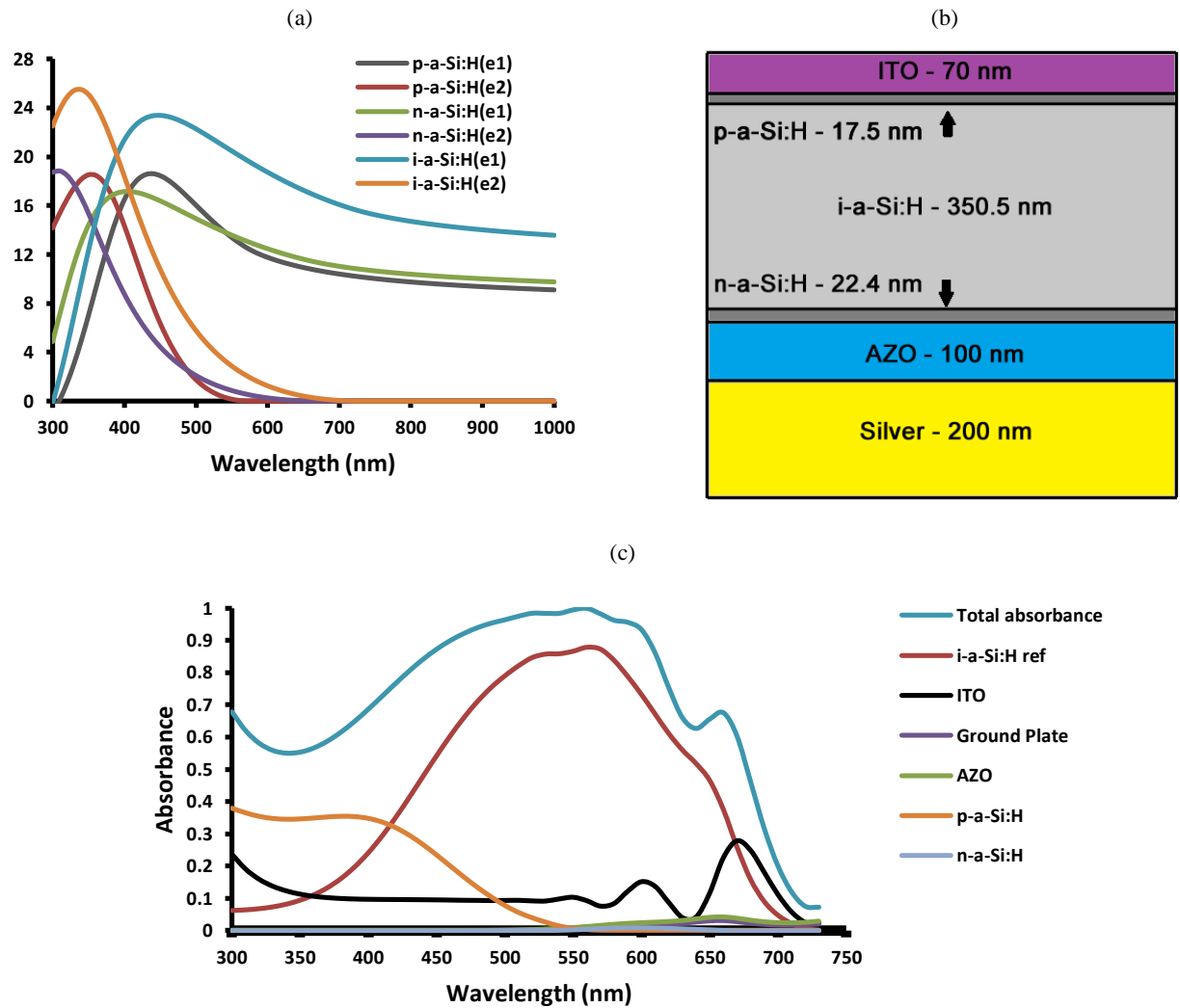

FIG. 1. (a) The measured value of complex permittivity for p, i and n a-Si:H layers using spectroscopic ellipsometry. (b) Reference structure, the top layer is ITO of 70 nm thickness, followed by p-a-Si:H of 17.5 nm thickness, then i-a-Si:H of 350.5 nm thickness, after that n-a-Si:H of 22.5 nm thickness, followed by AZO of 100 nm thickness and the last is a silver ground plate of 200 nm. (c) The plot for absorbance against wavelength for all the layers in reference structure.

The commercial reference structure in Fig. 1(b) is designed such that the stack of semiconductor and dielectric layers sandwiched between transparent conductive oxide (TCO) and back silver reflector supports broad Fabry-Perot-like resonance[32, 33] to maximize the absorption in the i-a-Si:H layer. The integral in equation (1) for this reference structure is evaluated to be 268.9 W/m$^2$ which is the absorbed power density in the i-a-Si:H layer of the reference for AM 1.5 spectra. This reference structure is strongly absorbing in the 500-600 nm region of the solar spectrum; however, the absorption in the i-a-Si:H layer falls sharply for wavelength greater than 600 nm. The parasitic losses in the ITO buffer increase significantly at blue ($\lambda$ <350 nm) and red ($\lambda$ >600 nm) regions of solar spectrum, resulting in up to 38.2 % of the total absorbance at 660 nm. The increased parasitic loses at blue (wavelength < 350 nm) region of solar spectrum in the ITO can be accounted by substantial increase in the attenuation coefficient of ITO,[29] mainly due to approaching band edge of ITO, causing escalated parasitic losses for blue spectrum. Whereas, at red (wavelength >600 nm) regions of solar spectrum, observed absorbance peaks in the ITO mainly result from reflections from multiple layers of the reference structure. When the electromagnetic field reflected from the top interface of the i-a-Si:H layer constructively interferes with the field incident from inside the i-a-Si:H layer, as a resulf of reflection from the bottom interface of the i-a-Si:H layer, absorbance peaks occur. On the other hand, if these waves destructively interfere, dips in the absorbance spectra appear.

It is evident from Fig. 2(b) that the NDPSC outperforms the reference for all parts of the solar spectrum, especially in the region of 550-720 nm because of a multi-resonance response. The absorbed power density in the i-a-Si:H layer of NDPSC calculated using equation (1) is 318.7 W/m$^2$, which is approximately 18.5 % higher than the reference. At normal incidence, the NDPSC is found to have a $J_{SC}$ of 14.0 mA/cm$^2$ as compared to 11.7 mA/cm$^2$ for the reference cell, which is a 19.7 % enhancement in $J_{SC}$.

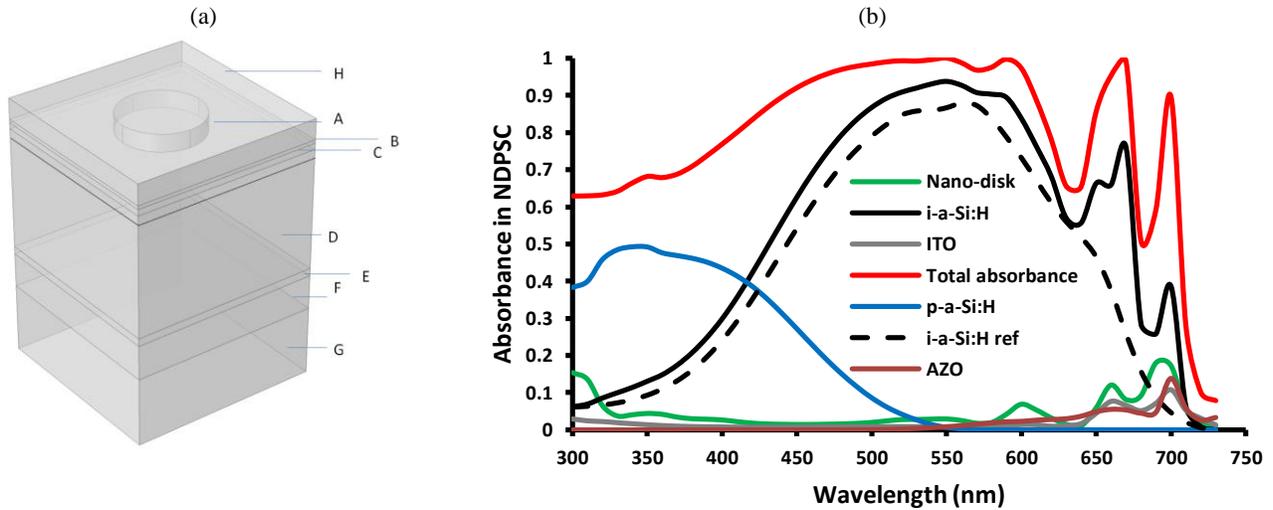

FIG. 2. (a) Nano-disk patterned a-Si:H thin-film solar cell A) nano-disk, diameter: 240 nm, height: 50 nm B) ITO layer, height: 10 nm, C) p-layer, height: 17.5 nm, D) i-layer, height: 350.5 nm, E) n-layer, height: 22.5 nm, F) AZO layer, height: 100 nm, G) silver layer, height: 200 nm, H) ARC (silicon nitride), height: 60 nm. Silver nano-disk is embedded in the silicon nitride ARC. The boundary of the top section is shown by black line. (b) The plot for absorbance against wavelength for all the layers in NDPSC, the absorbance in i-a-Si:H layer of reference structure is shown in a dashed line for comparison and very low absorbance in n-a-Si:H layer is not shown.

Optical absorption in NDPSC and reference is a function of incidence angle (from normal) and polarization; therefore $J_{SC}$ depends on incidence angle and polarization and has been illustrated in Fig 3. The $J_{SC}$ in the NDPSC is always higher than the reference structure for both the Transverse Electric (TE) and the Transverse Magnetic (TM) polarizations (except a negligible difference near 80°), and $J_{sc}$ for both TE and TM polarizations was found to be similar for all incidence angles;

making the NDPSC structure almost polarization independent to the incident light. For the reference structure, the $J_{sc}$ for TE and TM polarizations was found to be similar for all incidence angles, except around 60° where $J_{sc}$ was found to be ~5% higher for the TM polarization as compared to the TE polarization. However, $J_{sc}$ is not truly incidence angle independent and it was found that the $J_{SC}$ in both NDPSC and reference cells, for both TE and TM polarizations decreased gradually with increasing angle of incidence, and both plunge sharply after 60° and nearly converge at 80°; producing $J_{SC}$ of ~6.5 mA/cm$^2$.

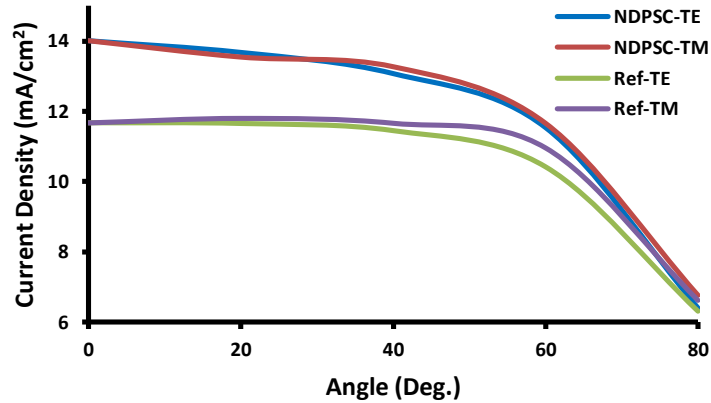

FIG. 3. Plot for the short circuit current density in NDPSC and reference solar cell, for both TE and TM polarizations as a function of angle of incidence from normal. The short circuit current density in the NDPSC for both TE and TM polarizations is similar and always higher than the short circuit current density of reference solar cell at all angles (except a negligible difference near 80°) and it converges to that of reference cell at 80°.

It has been reported that increasing the thickness of the metal patterned for plasmonic photovoltaic applications can significantly increase reflectance from top surfaces and increase Ohmic losses in the metal in the form of heat.[21, 34] However, it was found that when the height of the nano-disk patterned structure was systematically increased from 10 nm to 50 nm, the OE of NDPSC increased from 7.9 % to the maximum of 18.5% at 50 nm thickness. This can be attributed to improved impedance matching to the impedance of free space achieved by the top section (ARC and NDS combined) as a result of reduced geometric skin depth with increasing nano-disk thickness.[12, 35, 36] It is important to note from Fig. 2(b), that the nano-disk patterned structure improves the performance for a broad part of frequency spectrum, but itself becomes more lossy at shorter wavelengths (λ < 400 nm). At shorter wavelengths, losses in both nano-disk and p-a-Si:H layer increases significantly. At these shorter wavelengths, the p-a-Si:H layer account for major absorption instead of in the i-a-Si:H; the resultant recombination losses are dissipated as heat. This is the major drawback for ultrathin broadband a-Si:H photovoltaic cells, evident from Fig. 2(b), especially when the plasmonic patterned structure improves overall absorption in the semiconductor region; however, quite a significant portion of that absorption is shared by the ITO, p- and n-a-Si:H layers, producing heat by recombination.

In order to explore the possibility of broadband response with the current commercial solar cell structure, we simulated the p-layer having the same thickness as in commercial structures, 17.5 nm, with periodic boundaries on vertical sides and perfectly matched ports for top and bottom surfaces of the p-layer as shown in Fig. 4(a). Its response is shown in Fig. 4(b). It can be observed that the p-layer is not lossy for 550-730 nm, however for 300-550 nm, the p-layer becomes very lossy and accounts for major absorption at lower wavelengths near 300 nm. Therefore, it can be concluded that with the current p-i-n or n-i-p design with a p-layer on top of the active absorbing layer with respect to the solar source, broadband absorption for 300-730 nm is not achievable.

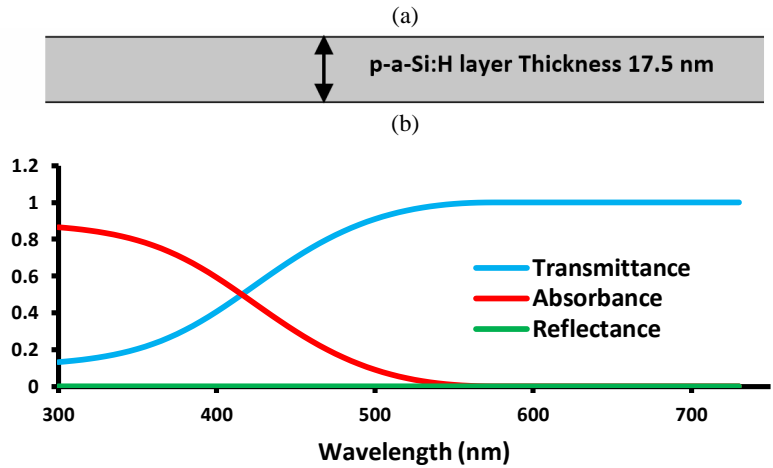

FIG. 4. (a) The p-a-Si:H layer of NDPSC. (b) Absorbance (A), Reflectance (R) and Transmittance (T) of the p-a-Si:H layer having matched ports (top and bottom), for thickness 17.5 nm.

The optical response of NDPSC has several resonant peaks which need to be analyzed to understand the physical mechanism responsible for such an enhancement. We analyzed the field distribution (Fig. 5) at all the peaks in optical response of NDPSC, i.e. for wavelengths at 550 nm, 590 nm, 650 nm, 670 nm and 700 nm. This analysis reveals the presence of Fabry-Perot like resonance inside the i-a-Si:H layer, which is similar to the field distribution as reported by Varadan, et al.[37] To better understand the nature of the resonance, the NDPSC in Fig. 2(a) was modeled as a Fabry-Perot resonator, having i-a-Si:H as a cavity medium. The section comprising the n-layer, AZO and ground plate were modeled as a bottom reflector. The top section comprising the p-layer, ARC, ITO and nano-disk patterned structure were modeled as a lossy top reflector. For qualitative analysis, we considered the simple equation for Fabry-Perot model, describing the resonance condition in such a cavity by:

$$L = \frac{m\lambda}{2\sqrt{\frac{\varepsilon'}{2} \cdot \left(\sqrt{1+\left(\frac{\varepsilon''}{\varepsilon'}\right)^2}+1\right)^{1/2}}} \qquad (4)$$

where L is the length of the cavity, λ is the wavelength in free space, $\varepsilon'$ and $\varepsilon''$ are the real and imaginary coefficients of complex relative permittivity (of the form $\varepsilon_r = \varepsilon' - j\varepsilon''$) of the cavity, and m is an integer. When the peaks of NDPSC were analyzed using equation (4) for optical response of the NDPSC at 550 nm, 590 nm, 650 nm, 670 nm and 700 nm wavelengths, the calculated cavity length was found to be 363.8 nm, 337.6 nm, 382.1 nm, 327.4 nm and 349.1 nm respectively. For a 700 nm wavelength, a cavity length of 349.1 nm resulted in good agreement with the thickness of the i-a-Si:H layer as 350.5 nm. For other wavelengths, the results are convincingly close to the thickness of i-a-Si:H layer. At other wavelengths, the small difference in the cavity length can be attributed to (i) the indistinguishable boundaries between the mirror and cavity medium due to the presence of inhomogeneous structure in top reflector/mirror and (ii) infringing near field produced by surface plasmons in the vicinity of metallic nano-disk patterned structure. At shorter wavelengths (λ<550 nm), the Fabry-Perot resonance is not sustainable for large cavity lengths because of very large value of $\varepsilon''$ of i-a-Si:H. This is confirmed from the absorbance in n-a-Si:H layer, which results in less than 0.0001% of total absorbance in spite of having similarly large value of imaginary part of permittivity as i-a-Si:H layer. This indicates that the incident solar radiation is not able to reach the n-a-Si:H layer, and is being absorbed before that or reflected. We should note that substantially large $\varepsilon''$ can also lead to poor impedance matching, which may be difficult to improve.[36, 38] This qualitative analysis is simple, without exploring the detailed parameter space, yet convincingly close to the results obtained by numerical computation using COMSOL in our work, and is similar to the method employed by Massiot, et al.[32]

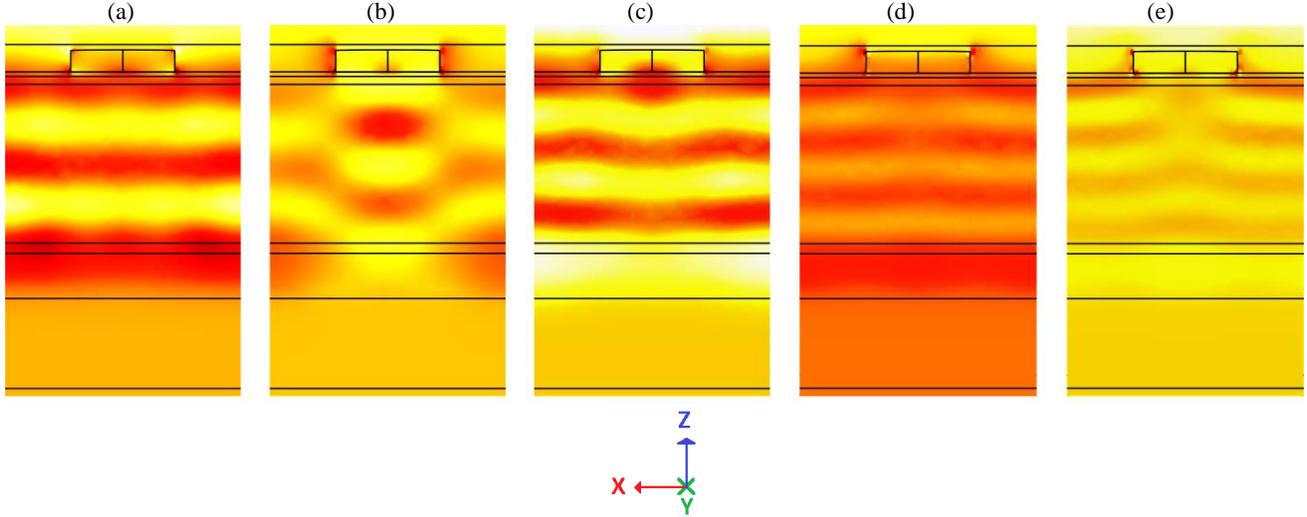

FIG. 5. $E_x$ - '$x$' component of electric field in NDPSC at resonance wavelengths of (a) 700 nm, (b) 670 nm, (c) 650 nm, (d) 590 nm and (e) 550 nm. The NDPSC is excited by wave having electric field, **E**, in the $x$ direction, wave vector, **k**, along negative $z$ direction, and magnetic field, **H,** along the y direction. The field distribution resembles Fabry-Perot resonance in the i-a-Si:H layer as a cavity medium. The Fabry-Perot resonance is responsible for major enhancement at all the resonance peaks.

To further understand the reason behind the excitation of multiple Fabry-Perot resonances in the NDPSC, the top section of NDPSC as portrayed in Fig. 2(a) (comprising of the ARC, nano-disk, ITO, p-a-Si:H layer and 5 nm of i-a-Si:H layer) was carefully investigated. In Fig. 2(b), it can be noted that there are few prominent metallic absorbance peaks around 600nm, 660nm, and 700 nm suggesting field enhancements due to metallic nano-disk structure (NDS). In order to further investigate the cause of these sharp absorbance peaks produced by the nano-disk patterned structure, we analyzed the current density distribution at the absorbance peaks near the nano-disk surfaces, as shown in Fig. 6. These resonances, which can be identified by their field distribution (see Fig. 6) in the top section of NDPSC, are responsible for the excitation of Fabry-Perot resonances. The significance of the existing plasmonic resonances in the top section residing at the metallic surfaces is to aid in design consideration and tuning of Fabry-Perot resonance around that region, such that the NDPSC can be modeled as a Fabry-Perot resonator with a i-a-Si:H layer sandwiched between the bottom mirror and the top section with reflectivity and surface impedance tuned by excited plasmonic resonances at the metallic surfaces. Therefore, at wavelengths of 550 nm, 590 nm, 650 nm, 670 nm and 700 nm in the absorbance spectra, the top section of the NDPSC aids in a strong resonance condition leading to almost perfect total absorbance by matching the surface impedance with that of free space. The nano-disk structure creates this impedance matched condition through excitation of surface plasmons between the Ag/ITO and Ag/$Si_3N_4$ interfaces. On the basis of Fig. 7, it can be observed that when there is a resonance-like condition in an Ag nano-disk, the reflectance from NDPSC is reduced to zero, creating a perfect total absorbance condition in which a major fraction of the total absorption is shared by the i-a-Si:H layer. This confirms that the nano-disk aids in the impedance matching of the surface impedance of NDPSC with the impedance of free space. This impedance matching condition owing to excitation of surface plasmons due to nano-disk helps in exciting a Fabry-Perot resonance in NDPSC where the i-a-Si:H layer acting as a cavity media. A similar mechanism has been recently studied in detail in the context of metamaterial perfect absorbers using a perfectly impedance matched sheet,[39] interference theory,[40-42] and grating theory[43] models.

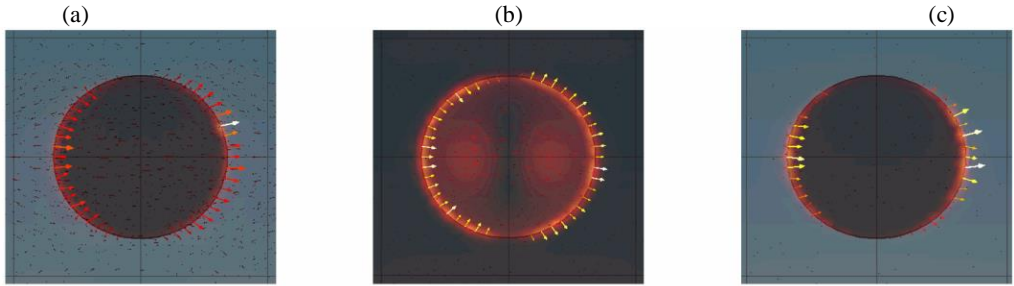

FIG. 6. Current density distribution near the nano-disk surfaces at resonance wavelengths of (a) 700 nm, (b) 660 nm, and (c) 600 nm. The arrows and color show the direction and magnitude of the current density in the given planes. At 700 nm and 600 nm the majority of the surface plasmon polaritons are excited at the Ag/Si$_3$N$_4$ interfaces while at 660 nm the majority resides at Ag/ITO interface. The planes shown correspond to the planes where the majority of the surface plasmon polaritons reside.

It can be noted that in the NDPSC structure the surface plasmons produced at the interface of Ag/ITO also contribute to some enhancement, especially around 650 nm, 670 nm and 700 nm as evident from the field distribution ($E_z$ - '$z$' component of electric field) in Fig. 8. The near field enhancement due to surface plasmons produced at Ag/ITO interface extends from the ITO and p-a-Si:H layer to the i-a-Si:H layer causing enhanced absorption especially at 650 nm, 670 nm and 700 nm. The near-field absorption is stronger in i-a-Si:H layer as compared to the ITO and p-doped layer, which have low $\varepsilon''$ and extremely small thickness compared to the i-layer. In most plasmonic solar cell literature[12-16, 19, 21, 22-23, 33] it is mainly this near field effect that is thought to be responsible from optical absorption enhancement. In contrast, we employ the plasmonic effect primarily for impedance tuning to excite strong Fabry-Perot resonances[32, 37] in the i-layer. It can be noted here that the thin ITO buffer layer aids in tuning the resonance towards a blue shift as also reported by Massiot, et al.[32] For shorter wavelengths, the absorption in the i-layer appears to be dominated by the large absorption coefficient of a-Si:H instead of a plasmonic enhancement or Fabry-Perot resonance.

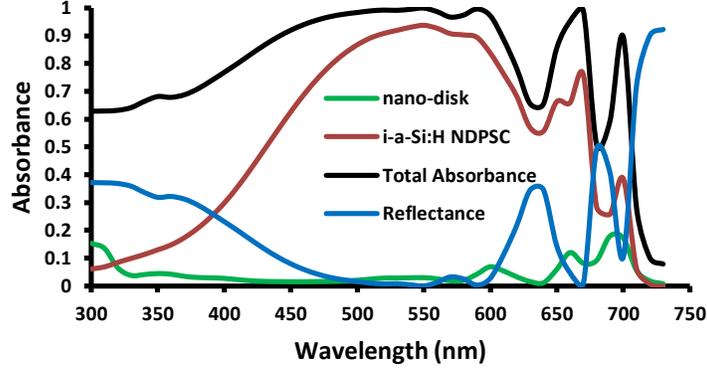

FIG. 7. When there is resonance condition in Ag nano-disk, indicating the excitation of surface plasmons at Ag/ITO interface, there is an impedance matching condition with free space and therefore reflectance from the NDPSC almost drops to zero and aids in enhanced absorption in the i-a-Si:H layer.

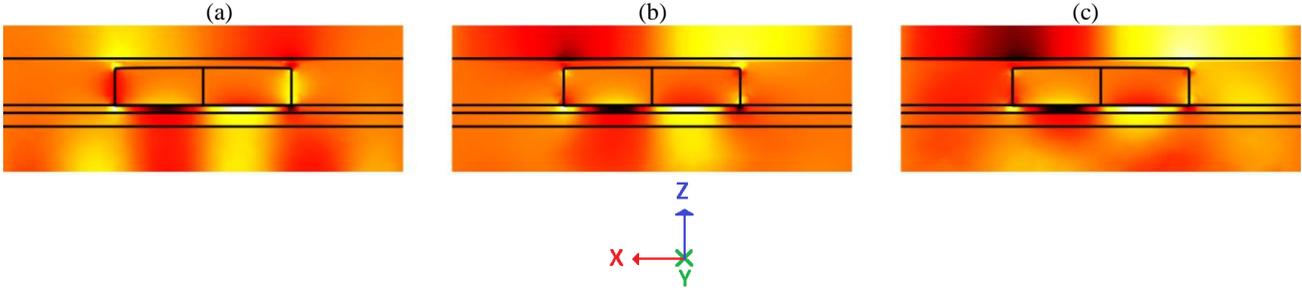

FIG. 8. $E_z$ field distribution in NDPSC at resonance wavelengths of (a) 700 nm, (b) 670 nm and (c) 650 nm. The NDPSC is excited by wave having electric field, **E**, in the *x* direction, wave vector, **k**, along negative *z* direction, and magnetic field, **H**, along the y direction. The strong field distribution just below the Ag disk indicates surface plasmon resonance at Ag/ITO interface producing near field penetrating into the i-a-Si:H layer responsible for partial enhancement and also impedance matching with free space and aiding in excitation of Fabry-Perot resonance.

We also simulated the effect of increasing the period of the unit cell (up to 1.5 μm) while keeping the other parameters fixed, to observe its effect on absorbed power density in the i-a-Si:H layer. It was found that the absorbed power density in the i-a-Si:H layer for the AM1.5 spectrum reached a maximum at a 550 nm period, and then exponentially reduced to the bulk value with no metallic patterned nano-disk, shown in Fig. 9. This result validates the argument that the effect of optical enhancement in the i-a-Si:H layer due to combined mechanisms of near field plasmonic enhancement and Fabry-Perot resonance is due to nano-disk metallic patterned structure.

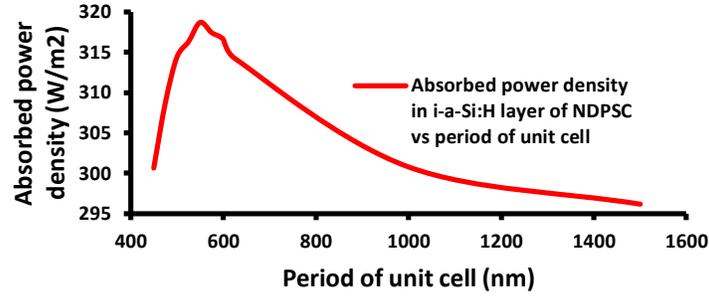

FIG. 9. The plot for absorbed power density in i-a-Si:H layer of NDPSC versus period of unit cell. The absorbed power density for AM 1.5 spectrum is maximum at 550 nm period and it exponentially decreases with increasing period converging towards the bulk value without any metallic nanostructure.

The total absorbed power density (i.e., absorbed power density corresponding to whole structure, for a solar AM 1.5 spectrum of 300-2500 nm) was found to be higher for NDPSC (615.5 W/m$^2$) as compared to the reference cell (471.3 W/m$^2$). The difference between total absorbed power density and absorbed power density in the i-layer is the minimal power density converted into heat. Numerically, for the NDPSC cell this is equal to 296.8 W/m$^2$ and 202.4 W/m$^2$ for the reference, which is a substantial difference of at least 46.6%. It should be noted that this calculation of total absorbed power density does not account for wavelengths larger than 2500 nm and lower than 300 nm, and other loss mechanism like recombination, conduction losses, etc. With these additional wavelengths and other loss mechanisms accounted for, the absorbed power density that is converted into the heat in the case of the NDPSC cell is larger than the reference. We calculated the temperature of the reference structure to be 55.9 °C based on the report by Skoplaki, et al.,[44] subsequently the additional nearly 50% power density converted into heat for NDPSC is expected to increase the operating temperature of the NDPSC cell over the reference cell appreciably. It has been observed that SWE is reversible in nature and the efficiency of the a-Si:H solar cell can be returned to its initial state by heating the cell at 150 °C for 4 hours, which leads to an annealing of defect states.[4] However, the defect states can also anneal at lower temperatures over more extended time.[7, 45] Therefore, the heat produced in the cell during normal operation in the case of NDPSC, which is at least 46.6% more compared to the reference, may aid in reducing the detrimental effects of SWE significantly by faster self-annealing and provide an improved absorber for solar photovoltaic thermal (PVT) hybrid systems.

It should be noted that the concept of utilization of additional generated heat for SWE compensation and defects state annealing is applicable to all types of plasmonics based thin film a-Si:H solar cell, not just limited to NDPSC. Any type of plasmonics based thin film a-Si:H solar cell having OE more than its corresponding reference will also enhance total absorption, and hence additional heat will be generated. It should be noted that SWE defects can even be annealed at room temperature[7] and that any additional heat will increase the rate of defect state annealing. There are many literature reporting on plasmonics based thin film a-Si:H solar cell, however none of them have discussed SWE compensation and defects state annealing using additional generated heat through this approach. We are reporting this concept for the first time. It must be further noted that in this study, we have chosen a specific commercial design to demonstrate the concept of achieving higher optical absorption and SWE compensation, however, this concept of achieving higher optical absorption in active layer and subsequently higher OE and SWE compensation is valid for any type of commercial thin film a-Si:H solar cell.

To analyze the electrical characteristics of both the reference cell and NDPSC, their current density vs. bias voltage curves (J-V curves, Fig. 10) for the reference cell and NDPSC were obtained by numerical computations performed in COMSOL Multiphysics v4.3b for the system of equations described by Wang et al.[46] The short circuit current density for the reference cell was found to be 10.95 mA/cm$^2$ and 13.27 mA/cm$^2$ for NDPSC, which is almost an improvement of 21.2 %. It was found that the current density in NDPSC was higher than the reference cell for all values of bias voltage; however, the open circuit voltage was similar for both the reference cell and the NDPSC, nearly equal to 0.82 V. On the basis of Fig. 10, it is evident that the NDPSC outperforms the reference cell in current density characteristics. It should be noted that this short-circuit current density is different from and smaller than the maximum theoretical short-circuit current density discussed earlier (calculated using equation (3)), on the account of recombination and other loss mechanisms in calculations.

As the results have shown fabrication of NDPSCs appear to be a technically viable method of improving a-Si:H PV performance. However, extremely high-tolerance fabrication of geometrically acceptable nanostructures is non-trivial. Conventional techniques like e-beam lithography, x-ray lithography, ion beam lithography, nano-imprint lithography and extreme ultraviolet lithography although feasible are currently economically prohibitive for solar cell mass-manufacturing for areas of 1 meter square and larger. To overcome this challenge we recommend adapting a well-tested method for size-controllable nanostructure array formation using self-assembled polystyrene beads.[47-49] In this nanosphere lithography method polystyrene beads are applied to the surface of the PV and close packed through self-assembly. The size of the columns and the spacing is controlled by selection of the initial bead sizes and subsequent reactive ion etching. In this way the nano-disk pattern shown in Fig. 2a can be fabricated over large areas of thin-film solar cells at reasonable expenses.

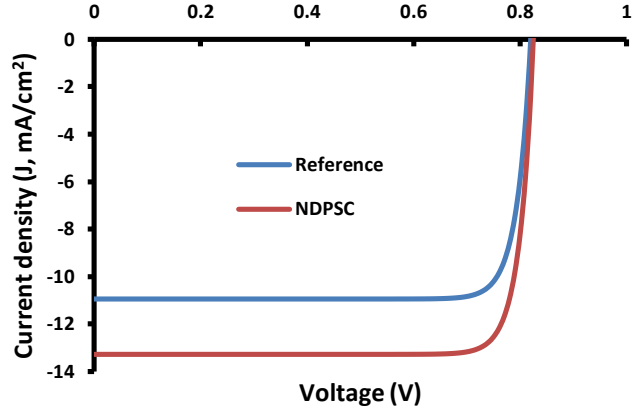

FIG. 10. The plot for current density versus bias voltage, for the reference cell and the NDPSC. The $V_{OC}$ was found to be 0.82 V for the reference and 0.825 V for the NDPSC. The short circuit current density was found to be 10.95 mA/cm$^2$ for the reference structure and 13.27 mA/cm$^2$ for the NDPSC.

To explore the impact of fabrication imperfections a detailed sensitivity analysis was also performed for a minimum of ±10% variation in the dimensions and parameters for the critical films and features, computed independently, and then in combinations. The change in optical enhancement was less than 2% of the optimum case for the parameters described in Fig. 2a.

## IV. CONCLUSIONS

In summary, we have discussed a front layer design for a-Si:H solar cell based on a multi-resonant plasmonic nano structure to significantly improve the optical performance of commercial a-Si:H solar photovoltaic cells and simultaneously combat the detrimental effects of SWE. The effects of absorption in the ITO, p and n-a-Si:H layers were considered and the absorption was maximized in i-a-Si:H layer for a commercial a-Si:H solar cell, without changing the parameters of semiconducting layers. We have numerically demonstrated the multi-resonant, polarization independent and optically enhanced response in a commercial cell design by introducing an array of metallic nano-disk patterned structures embedded in a silicon nitride anti-reflection coating that facilitate an absorption enhancement in i-a-Si:H layer over a significant (450 nm < λ < 720nm) fraction of the solar spectrum. Due to the axial symmetry of nano-disk structure, polarization independence is achieved. The total enhancement in optical absorption for AM 1.5 solar spectrum was found to be 18.5 % higher than the reference cell and exhibiting a modeled $J_{SC}$ of 14.0 mA/cm$^2$ compared to 11.7 mA/cm$^2$ for the reference, an improvement of 19.7 %. The NDPSC was also found to be superior in optical absorption at all angles of solar radiation incidence up to 80°. The $J_{sc}$ in NDPSC was also found to be similar for both TE and TM polarizations for all incidence angles from normal up to 80° when they finally merged with that of reference. We also demonstrated that it is not possible to get a broadband absorption response for the 300-550 nm part of spectrum for the p-i-n design due to the highly lossy p-a-Si:H layer. Therefore this multi-resonant response is so far the best response that can be achieved for conventionally designed and commercialized a-Si:H solar cells using plasmonic nanostructures. The substantial additional heat generated by the NDPSC can contribute to reducing SWE and improving the electrical characteristics of the a-Si:H commercial solar cell by self-annealing of defect states. The associated mechanism of optical enhancement in the NDPSC can be attributed to combined plasmonic and Fabry-Perot resonance by nano-disk structures.

**ACKNOWLEDGMENTS**

This work was supported by the National Science Foundation under grant award number CBET-1235750. We would like to thank Satyadhar Joshi, Arpit Ludhiyani and Jaya Jagannath Das for insightful discussions.